\documentclass{svproc}
\usepackage{graphicx}
\usepackage{amsmath}
\usepackage{cite}
\usepackage{booktabs}

\usepackage{url}

\begin{document}
\mainmatter              
\title{Two-nucleon emitters within a pseudostate approach}
%
%
\author{J. Casal\inst{1,2} \and J. G\'omez-Camacho\inst{3,4}}
%
%
%
\institute{European Centre for Theoretical Studies in Nuclear Physics and Related Areas (ECT$^*$), Strada delle Tabarelle 286, I-38123 Trento, Italy
\and
Dipartimento di Fisica e Astronomia ``G.~Galilei'' and INFN - Sezione di Padova, Via Marzolo 8, I-35131 Padova, Italy,\\
\email{casal@pd.infn.it}
\and
Departamento de F\'{\i}sica At\'omica, Molecular y Nuclear, Facultad de F\'{\i}sica, Universidad de Sevilla, Apartado 1065, E-41080 Sevilla, Spain
\and
Centro Nacional de Aceleradores (CNA), U.~Sevilla, J.~Andalucía, CSIC, Tomas A.~Edison 7, E-41092 Sevilla, Spain
}

\maketitle              

\begin{abstract}
A method to identify and characterize three-body resonances in a discrete basis is discussed in the context of two-nucleon emitters. For this purpose, a resonance operator is introduced and diagonalized in a basis of energy pseudostates within the hyperspherical formalism. Then, the energy and width of the resonance are obtained from its time dependence. The approach is illustrated for $^{16}$Be ($^{14}\text{Be}+n+n$).
\end{abstract}
\section{Introduction}
The study of two-nucleon correlations beyond the driplines has gained renewed attention since the recent experimental observation of direct two-proton~\cite{giovinazzo02,grigorenko09} and two-neutron~\cite{spyrou12,kohley13} decays. These are typically discussed in terms of different possible paths: The so-called sequential, direct and democratic decays~\cite{pfutzner12}. From the theoretical point of view, a proper description of three-body resonances can help in understanding these correlations~\cite{lovell17,JCasal18}. The description of few-body resonant states, however, is not an easy task. In this work, we briefly describe a robust approach to identify and characterize three-body resonances in a discrete basis within the hyperspherical formalism, and we apply the method to $^{16}$Be ($^{14}\text{Be}+n+n$). For further details, see Ref.~\cite{casal19}.

\section{Resonance operator}

In general, resonances correspond to a range of continuum energy eigenstates whose probabilities are concentrated within the potential well. Since these continuum structures will be very sensitive to changes in the potential, we introduce the resonance operator
\begin{equation}
    \widehat{M}=\widehat{H}^{-1/2}\widehat{V}\widehat{H}^{-1/2}, ~~~~ \widehat{M}|\psi\rangle = m |\psi\rangle
    \label{eq:resop}
\end{equation}
whose eigenstates $|\psi\rangle$ can be expanded in Hamiltonian pseudostates $|n\rangle$,
\begin{equation}
    |\psi\rangle = \sum_n \mathcal{C}_n|n\rangle, ~~~~\widehat{H}|n\rangle = \varepsilon_n |n\rangle.
    \label{eq:eigen}
\end{equation}
Note that, if the system has no bound states, all energies $\varepsilon$ are positive. Therefore, since $m\sim V/\varepsilon$, resonances can be identified from the eigenstates of $\widehat{M}$ corresponding to large negative eigenvalues.

\begin{figure}
\centering
\includegraphics[width=0.77\linewidth]{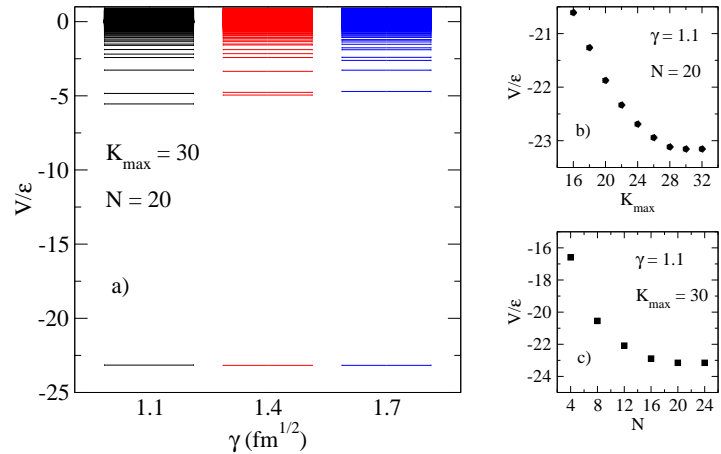}
\caption{a) Eigenvalues of $\widehat{M}$ for $^{16}$Be(0$^+$) as a function of the basis parameter $\gamma$. The right panel shows the convergence of the lowest eigenstate as a function of b) the maximum hypermomentum $K_{max}$ and c) the number of basis functions $N$. In each case, the other two parameters are fixed.}
\label{fig:gamma}
\end{figure}

We apply the method to study $^{16}$Be(0$^+$) states in a three-body ($^{14}\text{Be}+n+n$) model using the potentials in Refs.~\cite{lovell17,JCasal18}. Calculations are performed within the hyperspherical formalism~\cite{Zhukov93}, where the relevant parameters are the maximum hypermomentum $K_{max}$ (which determines the number of angular components in the wave-function expansion), the number of basis functions $N$, and the scale parameter $\gamma$ controlling the radial extension of the basis\cite{JCasal13}. Here, smaller $\gamma$ values correspond to more extended basis functions and a larger concentration of energy pseudostates just above the breakup threshold. In Fig.~\ref{fig:gamma}, we present the spectra of $\widehat{M}$ for three different bases, where the lowest eigenstate is stable and clearly separated from the rest. This state represents the ground-state resonance of $^{16}$Be, which shows a fast convergence with respect to the size of the model space, as shown in the right panel. As discussed in Ref~\cite{JCasal18,casal19}, the corresponding wave function presents a dominant dineutron component, which favors the picture of a correlated two-neutron emission from the ground state of $^{16}$Be.

\section{Time dependence and resonance parameters}

As time evolves, the state given by Eq.~(\ref{eq:eigen}) loses its character, and we can define a time-dependent amplitude
\begin{equation}
    |\psi(t)\rangle = \sum_n \mathcal{C}_n e^{-i\varepsilon_n t}|n\rangle, ~~~~ a(t)=\langle\psi(0)|\psi(t)\rangle = \sum_n|\mathcal{C}_n|^2 e^{-i\varepsilon_n t}.
    \label{eq:time}
\end{equation}
For an ideal Breit-Wigner resonance, we would expect
\begin{equation}
    a_r(t)=e^{-i\varepsilon_r t - \frac{\Gamma}{2} t},
    \label{eq:resamp}
\end{equation}
given by the resonance energy $\varepsilon_r$ and its width $\Gamma$. These parameters can be chosen so that Eq.~(\ref{eq:resamp}) is as close as possible to the amplitude in Eq.~(\ref{eq:time}). We define a resonance quality parameter with the meaning of a quadratic deviation
\begin{equation}
\delta^2(\varepsilon_r,\Gamma)=\frac{\int_0^\infty e^{-xt}\left|a(t)-a_r(t)\right|^2 dt}{\int_0^\infty e^{-xt}\left|a(t)\right|^2 dt},
\label{eq:rqp}
\end{equation}
where $1/x$ corresponds to a relevant time scale for the resonance formation. Thus, small values of $x$ will be related to long times associated to the decay. The resonance parameters, as a function of $x$, can be obtained by minimizing Eq.~(\ref{eq:rqp}),
\begin{equation}
\frac{\partial}{\partial \varepsilon_r}\delta^2(\varepsilon_r,\Gamma)=0, ~~~~ \frac{\partial}{\partial \Gamma}\delta^2(\varepsilon_r,\Gamma)=0.
\end{equation}
Details on the derivation of the relevant equations can be found in Ref.~\cite{casal19}.

\begin{figure}[h]
\centering
\includegraphics[width=0.94\linewidth]{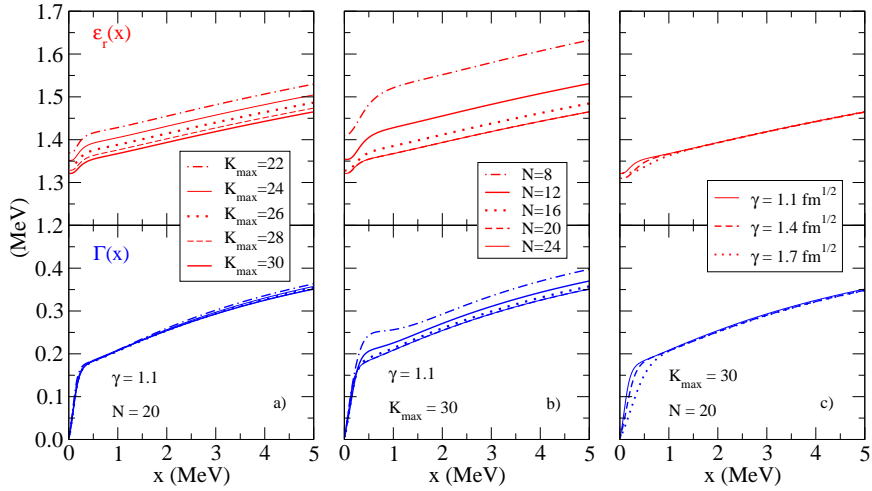}
\caption{Convergence of the resonance parameters $\varepsilon_r$ and $\Gamma$ with respect to a) $K_{max}$, b) $N$ and c) $\gamma$. In each case, the other two parameters are fixed.}
\label{fig:conv}
\end{figure}

For the $0^+$ ground state of $^{16}$Be, we obtain the results presented in Fig.~\ref{fig:conv}. Here we show the convergence of the resonance parameters $\varepsilon_r(x)$ and $\Gamma(x)$ with respect to $K_{max}$, $N$ and $\gamma$ as in the preceding section. With $K_{max}=30$ and $N=20$, we show that the resonance energy and width are fully converged. Both functions follow approximately a linear trend, with a small slope, for small values of $x$. Then, a sudden drop of the width is observed close to zero. As discussed in Ref.~\cite{casal19}, this occurs when a pseudostate energy $\varepsilon_n$ matches the resonance energy and is a consequence of the discrete nature of the basis. By increasing the level density near the threshold (i.e., choosing smaller values of $\gamma$), the linear trend explores smaller values of $x$. Therefore, it is reasonable to fix the resonance parameters $\varepsilon_r$ and $\Gamma$ by extrapolating this linear behavior to $x=0$. Following this prescription, for an energy of $\varepsilon_r(0^+)=1.341$ MeV we obtain $\Gamma(0^+)=0.160$ MeV. The convergence of these values with respect to $K_{max}$ and $N$ is shown in Table~\ref{tab:conv}. The computed width is consistent with the results in Ref.~\cite{lovell17} from the three-body eigenphases within the hyperspherical $R$-matrix method to solve the actual three-body scattering problem. This is an indication that the method here presented provides a reasonable description of three-body resonances in a discrete basis.

\begin{table}[ht]
    \vspace{-0.3cm}
    \centering
    \begin{tabular}{ccccccc}
    \toprule
    $K_{max}$ & $\varepsilon_r$ (MeV) & $\Gamma$ (MeV)  & \hspace{1cm} & $N$ & $\varepsilon_r$ (MeV)  & $\Gamma$ (MeV)  \\
    \midrule
    22 & 1.403 & 0.163 & & 8  & 1.489 & 0.228\\
    24 & 1.379 & 0.160 & & 12 & 1.399 & 0.188\\
    26 & 1.363 & 0.159 & & 16 & 1.359 & 0.168\\
    28 & 1.347 & 0.159 & & 20 & 1.342 & 0.160 \\
    30 & 1.341 & 0.160 & & 24 & 1.341 & 0.160 \\
    \bottomrule
    \end{tabular}
    \caption{Convergence of $\varepsilon_r$ and $\Gamma$ as a function of $K_{max}$ and $N$, with $\gamma=1.1$ fm$^{1/2}$.}
    \label{tab:conv}
    \vspace{-0.8cm}
\end{table}

\vspace{-5pt}

\section*{Acknowledgements}
This work has been supported by the Spanish Ministerio de Ciencia, Innovaci\'on y Universidades  and FEDER funds under Projects  No.~FIS2017-88410-P, FPA2016-77689-C2-1-R and FIS2014-51941-P, and by the European Union's Horizon 2020 research and innovation program under grant agreement No.~654002.

%

%
%
%
\bibliographystyle{spmpsci}
\bibliography{bibfile}

\begin{thebibliography}{10}
\providecommand{\url}[1]{{#1}}
\providecommand{\urlprefix}{URL }
\expandafter\ifx\csname urlstyle\endcsname\relax
  \providecommand{\doi}[1]{DOI~\discretionary{}{}{}#1}\else
  \providecommand{\doi}{DOI~\discretionary{}{}{}\begingroup
  \urlstyle{rm}\Url}\fi

\bibitem{giovinazzo02}
Giovinazzo, J., et~al.: Phys. Rev. Lett. \textbf{89}, 102501 (2002)

\bibitem{grigorenko09}
Grigorenko, L.V., et~al.: Phys. Lett. B \textbf{677}, 30 (2009)

\bibitem{spyrou12}
Spyrou, A., et~al.: Phys. Rev. Lett. \textbf{108}, 102501 (2012)

\bibitem{kohley13}
Kohley, Z., et~al.: Phys. Rev. Lett. \textbf{110}, 152501 (2013)

\bibitem{pfutzner12}
Pf\"utzner, M., et~al.: Rev. Mod. Phys. \textbf{84}, 567 (2012)

\bibitem{lovell17}
Lovell, A.E., Nunes, F.M., Thompson, I.J.: Phys. Rev. C \textbf{95}, 034605
  (2017)

\bibitem{JCasal18}
Casal, J.: Phys. Rev. C \textbf{97}, 034613 (2018)

\bibitem{casal19}
Casal, J., G\'omez-Camacho, J.: Phys. Rev. C \textbf{99}, 014604 (2019)

\bibitem{Zhukov93}
Zhukov, M.V., et~al.: Phys. Rep. \textbf{231}, 151 (1993)

\bibitem{JCasal13}
Casal, J., Rodr\'{\i}guez-Gallardo, M., Arias, J.M.: Phys. Rev. C \textbf{88},
  014327 (2013)

\end{thebibliography}
\end{document}